\documentclass[preprint,showpacs,preprintnumbers,amsmath,amssymb,prb,superscriptaddress]{revtex4-1}

\usepackage[usenames,dvipsnames]{color}
\usepackage{graphicx}
\usepackage{dcolumn}
\usepackage{simplewick}
\usepackage{bm}
\usepackage{amsmath}
\usepackage{float}
\usepackage{tabularray}
\usepackage[basic,lgr,defaultmathsizes]{mathastext}
\usepackage{wrapfig}

\usepackage{threeparttable, tablefootnote}
\usepackage{hyperref}
\usepackage{natbib}
\usepackage{hyperref}
\hypersetup{colorlinks=true, urlcolor=blue, citecolor=blue}

\newcommand{\cgb}{{\rm CsGeBr$_3$}}

\begin{document}

\title{Ferroelectric phases and phase transitions in CsGeBr$_3$ induced by mechanical load }

\author{Joshua Townsend}
\email{jdtownsend@usf.edu}
\affiliation{Department of Physics, University of South Florida, Tampa, Florida 33620, USA}
\author{Ravi Kashikar}
\affiliation{Department of Physics, University of South Florida, Tampa, Florida 33620, USA}
\author{S. Lisenkov}
\author{I. Ponomareva}
\email{iponomar@usf.edu}
\affiliation{Department of Physics, University of South Florida, Tampa, Florida 33620, USA}

\date{\today}

\begin{abstract}
First-principles-based atomistic simulations are used to reveal ferroelectric phases and phase transitions induced in a semiconductor ferroelectric, \cgb, by external loads: hydrostatic pressure, uniaxial and biaxial stresses, and misfit strain. Hydrostatic pressure was found to suppress the Curie point at the rate -0.45$T_C(0)$ K/GPa, where $T_C(0)$ is the zero pressure Curie temperature. Stresses and misfit strains were found to induce additional ferroelectric phase transitions and  phases not available under normal conditions.  We find that tensile load significantly enhances both the Curie temperature and spontaneous polarization, while compressive load has the opposite effect but  with the difference  that the Curie temperature is only slightly suppressed. The isothermal dependencies of polarization on pressure and stresses are highly nonlinear, which could result in large nonlinear piezoelectric responses. The phase diagrams reveal the diversity of the phases accessible through mechanical load, which include tetragonal, orthorhombic and monoclinic symmetries in addition to the rhombohedral and cubic ones realizable under normal conditions. We believe that this work reveals  the potential of Ge-based halide perovskites for applications in energy converting devices, which is especially significant in the current pursuit of environmental friendly lead-free technologies.

\end{abstract}
\maketitle

\section{Introduction}

Ferroelectric materials exhibit polarization in the absence of an electric field. This spontaneous polarization typically appears  below a certain temperature, known as the Curie point.
Ferroelectrics gained a significant attention in the 1940s with the discovery of a group of exceptionally robust ferroelectrics among oxide perovskites, compounds with a chemical formula ABO$_3$. The associated ferroelectric phase transitions in these materials give rise to many scientifically appealing and technologically attractive properties, such as extraordinary large dielectric constants reaching 10,000 in NBCTO\cite{dielectric1,dielectric2}, piezoelectricity ($d_{33} > \ 2500 \ pC/N$ in PZN-8\% PT\cite{highpiezo}),  pyroelectricity (up to 3000 to 5000 $\mu C \ m^{-2} \ K^{-1}$ in PMN-PT\cite{LargePyro}) , and nonlinear optical properties\cite{ReviewOxidePerovskites}. 
The diverse properties exhibited by oxide ferroelectrics have opened the door to many applications, such as mechanical sensors, actuators, temperature sensing devices, and nonvolatile computer memory. It is, however, uncommon to find semiconductor ferroelectrics among oxide perovskites. For oxide perovskites, this is seen due to a large difference in electronegativity between the metal cation (B) and the oxygen (O)\cite{bandgap2,bandgap}.  Very recently, inorganic lead-free halide perovskite family  CsGeX$_3$ (X=Cl, Br, I) has demonstrated ferroelectricity\cite{FerroHalides}, as well as many desirable optoelectronic properties, such as highly tunable bandgaps in the visible light regime\cite{TunableBandgap,TunableOpt2}, nonlinear optical properties\cite{NonlinearOpt}, and high photocurrent\cite{highphotocurrent,OpticalGeneral}, to name a few. Unlike oxide ferroelectrics, which possess large band gaps, the band gaps of the inorganic halide perovskites lie between 1.6 to 3.3 eV\cite{FerroPerovskites}, making them suitable candidates for solar cell applications, detectors, LED devices, and field effect transistors. Furthermore, this particular group of inorganic halide perovskites may be able to provide a lead-free alternative to many applications featuring lead-based perovskites, providing a path toward environmentally friendly technologies. 

This family of materials has been reported to undergo a single phase transition from cubic Pm$\bar{3}$m phase to rhombohedral R3m phase at temperatures of 428~K, 511~K, and 550~K for CsGeCl$_3$, CsGeBr$_3$, and CsGeI$_3$, respectively\cite{thiele}. Recent experimental studies have shown the existence of ferroelectricity in CsGeX$_3$ (X = Cl, Br, I) with spontaneous polarization of 12-15~$\mu$C/cm$^2$ for X=Br and 20~$\mu$C/cm$^2$ for X=I\cite{FerroPerovskites}. Currently, no experimental reports have been produced for polarization for X=Cl. The computational study Ref.\onlinecite{PolarAntipolar} has proposed that a dynamical antipolar phase may overlap with the cubic phase above the Curie temperature. One of the exceptional features of ferroelectrics is that they are also piezoelectrics, often times excellent ones. This suggests that application of mechanical load may have a profound effect on the ferroelectric phases and phase transitions in ferroelectrics. Indeed, we find multiple examples of phase tunability by mechanical load among ferroelectric oxides\cite{Phenomenal1,ExpPress1,CompPress1,presstune,straintune,straintune2}. At the same time virtually nothing is known about the piezoelectricity and associated electro-mechanical couplings in Ge-based halide ferroelectrics. How does the mechanical load affect ferroelectric phase transitions in such materials? Application of mechanical load is known to break the crystal symmetry, and therefore, is capable of inducing novel phases, which often times may have lower symmetry and very different properties. Lower symmetry phases may cause polarization rotation which in turn is capable of enhancing piezoelectric response \cite{BFOpolarrotate}. Moreover, the Curie point of stressed materials could differ significantly from the Curie point of unstressed materials and may open a way to further property tunability through Curie temperature engineering. It should be noted that very little is known about the response of Ge-based perovskites to mechanical load at present. Structural phase transitions in \cgb\ were reported in response to hydrostatic pressure, effectively lowering the Curie temperature\cite{ExpPress1}. The bandgap could also be manipulated by the application of hydrostatic pressure\cite{CompPress1}. 

Computationally, addressing the aforementioned questions is possible through application of finite-temperature methodologies that have been developed for ferroelectrics. The examples are phase-field methods\cite{PhaseField1}, phenomenological thermodynamic theories\cite{Phenomenal1}, first-principles-based effective Hamiltonians\cite{Vanderbilt1,Vanderbilt2} and bond-order methods\cite{BondOrder}. It should be mentioned that Density Functional Theory (DFT) is presently not used for finite-temperature simulations of ferroelctrics due to its extreme computational cost. On the other hand, the effective Hamiltonian approach \cite{Vanderbilt1,Vanderbilt2} has been specifically developed to extend the reach of DFT simulations of ferroelectrics to finite temperatures. Recently, such an approach has been developed for \cgb \cite{PolarAntipolar}, which allows us to pursue the investigation of ferroelectricity in this material under mechanical load.

The aim of this work is to use a multiscale approach that combines both effective Hamiltonian and DFT simulations in order to: (i) predict stress/strain induced stabilization of novel ferroelectric phases in CsGeBr$_3$, none of which are accessible under unstressed conditions;  (ii) report  rich and unusual set of phase diagrams in this material; (iii) to quantify property enhancement and tunability in this material by mechanical load, which feature up to a 35.9\% increase in Curie point and up to a 5.5\% increase in polarization under 1.0~GPa uniaxial tensile stress.

\section{Computational Methodology}
To achieve our goals we take advantage of the effective Hamiltonian approach, which was originally developed for oxide perovskites \cite{Vanderbilt1,Vanderbilt2,ReviewOxidePerovskites} and extended for halides in Ref.\onlinecite{PolarAntipolar}. In such an approach only the degrees of freedom that are relevant for ferroelectricity are retained. These are the local modes, $\mathbf u_i$, which are proportional to the local dipole moment of the perovskite unit cell, and both homogeneous and inhomogenoeus strain variables, ${\eta}$. The collective motion of local modes describe the soft zone-center ferroelectric mode. Strain variables describe long-wavelength acoustic phonons. Each unit cell contains one local mode vector and one inhomogeneous strain tensor, which allows capturing complex atomistic orderings of electric dipoles, such as vortices, nanodomains, and skyrmions\cite{PZTdomains,nanodomains,FEdomains}.

The homogeneous strain variables allow for supercell deformation as a whole, while inhomogeneous strain variables allow for the capture of unit cell deformations. The energy of the effective Hamiltonian is
\begin{equation}
  E^{total}  =   E^{self}( \{ \mathbf{u_i} \} ) \ + \ E^{dpl}( \{ \mathbf{u_i} \} ) \ + \ E^{short}( \{ \mathbf{u_i} \} ) \  + \ E^{elas}( \{ \mathbf{\eta_i} \} ) \ + \ E^{int}( \{ \mathbf{u_i} \} , \{ \mathbf{\eta_i} \}) \ - \ \sigma_j\eta_j \ + \ PV   
  \label{eq:1}
\end{equation}

and includes local mode self energy, $E^{self}$ (harmonic and anharmonic contributions), long-range dipole-dipole interaction, $E^{dpl}$, short-range interaction between local modes, $E^{short}$, elastic energy, $E^{elas}$, the interaction between the local modes and strains, $E^{int}$, and the energies that couple external stress, strain, pressure, and volume. The effective Hamiltonian has been previously used to model ferroelectric phase transitions in oxide ferroelectrics \cite{ReviewOxidePerovskites,Vanderbilt1, Vanderbilt2, RabePT,BCOphase}, and their tunability by the mechanical boundary conditions \cite{Phenomenal1, BTOstrain, McCash,nanomech}. In Ref. \onlinecite{PolarAntipolar} the effective Hamiltonian was used to study the ferroelectric phase transition in CsGeBr$_3$. With r$^2$SCAN based parametrization, it predicted the transition from cubic Pm$\bar{3}$m phase to ferroelectric rhombohedral R3m phase at 340 K, which underestimates the experimental transition temperature of 511 K\cite{PolarAntipolar}. The underestimation may be caused by the overestimation of the energy of the rhombohedral phase by the chosen exchange-correlation functional. It has been proposed that the SCAN family functionals do not adequately correct delocalization errors\cite{SCANdelocal,HSEcgb}, which could be an issue for structures with expressed lone pairs, as the rhombohedral phase is in our case. Indeed, HSE increases the energy difference between the ground state and cubic phase by a factor of 1.4, which could bring transition temperatures predicted by effective Hamiltonian in agreement with experimental data. However, parametrization with higher-level functionals, like HSE, is presently computationally prohibitive. It was shown to reproduce experimental coercive field and polarization. In this study we use the effective Hamiltonian as parameterized from r$^2$SCAN in the framework of classical Molecular Dynamics (MD) simulations to model bulk CsGeBr$_3$  at finite temperatures and under a wide range of mechanical loads. We model a supercell of 30$\times$30$\times$30 unit cells of CsGeBr$_3$ repeated periodically along all three Cartesian directions to simulate bulk crystal. Newton's equations of motion are solved numerically using a predictor-corrector integrator\cite{UnderstandingSim,ArtOfSim} and an integration step of 0.5 fs. The temperature is controlled through the Evans-Hoover thermostat\cite{UnderstandingSim}. Equilibrium thermodynamical phases are harvested from annealing simulations, which are initiated at 700 K and proceeded in steps of 10 K down to 10 K. In the case of hydrostatic pressure we encountered T$_C$ higher than 700~K and those simulations were redone starting  at 800~K. For each temperature we simulated 300,000 MD steps, and used 100,000 steps for equilibration and 200,000 steps for collecting thermal averages. Annealing simulations have been carried out for each mechanical load investigated. We studied hydrostatic pressure, uniaxial and biaxial stresses, and misfit, or epitaxial strain. These are motivated by their experimental realizability and scientific appeal. For example, misfit strain is nearly ubiquitous for thin films and is a powerful tool to control ferroelectric properties\cite{Phenomenal1,ThinFilmStrain1,ThinFilmStrain2,ThinFilmStrain3,ThinFilmStrain4}.  Hydrostatic pressure is known to have a profound effect on ferroelectric phase transitions. It destabilizes the ferroelectric phase through weakening long range interactions relative to short range interactions\cite{LinesGlass} and therefore can be used to fine tune the Curie point. Uniaxial stress is realizable during piezoelectric measurements, while biaxial stress occurs in partially relaxed epitaxial thin films \cite{StressThinFilms}. Technically, we simulate misfit strain in the range -5.0\% to 5.0\% by constraining components of the strain tensor (in Voigt notations) as follows $\eta_1=\eta_2=\frac{a-a_0}{a_0}$ and $\eta_6 = 0$, where $a_0$ is the cubic lattice constant of CsGeBr$_3$. In the simulated strain interval we consider the points with 0.5\% stepsize. For each value of strain we carry out annealing simulations and harvest the equilibrium polarization vector and lattice constants. These data are used to analyze the space group of the structure using ISOTROPY software\cite{FINDSYM,FINDSYM2}. Some representative structures are then subjected to  structural relaxation  using DFT as implemented in VASP \cite{VASP1,VASP2}. In such calculations we keep in-plane lattice vectors fixed to the same values as in effective Hamiltonian calculations to maintain the same mechanical boundary conditions, while allowing the other lattice vectors to relax. This is combined with full ionic relaxations. 
For DFT calculations we use the projector-augmented wave (PAW) pseudopotentials\cite{PAW} with  the $r^2$SCAN exchange-correlation functional\cite{r2scan}, a 10~$\times$~10~$\times$~10 reciprocal space grid  (grid density \(\sim \)  $ 0.02~\mathring{A}^{-1} $), and an energy cutoff of 800~eV. For piezoelectric tensor calculations we used the finite difference method with self-consistent response to finite electric fields as implemented in VASP. 

Similarly, biaxial stress is simulated by constraining components of the stress tensor as follows $\sigma_1=\sigma_2=\sigma$ and $\sigma_6=0$, while all other components are allowed to relax. We simulated the values of $\sigma$ in the range -2.0 GPa to 2.0 GPa with an interval of 0.5 GPa.  Uniaxial stress was modeled by fixing a single component of the stress tensor to a value from the range -2.0 GPa to 2.0 GPa with an interval of 0.5 GPa. Likewise, hydrostatic pressure was simulated in the range -2.0 GPa to 2.0 GPa, with an interval of 0.5 GPa.

\section{Phase diagrams predicted from finite-temperature simulations}

\begin{figure}[h]
\centering
\includegraphics[width=1\linewidth]{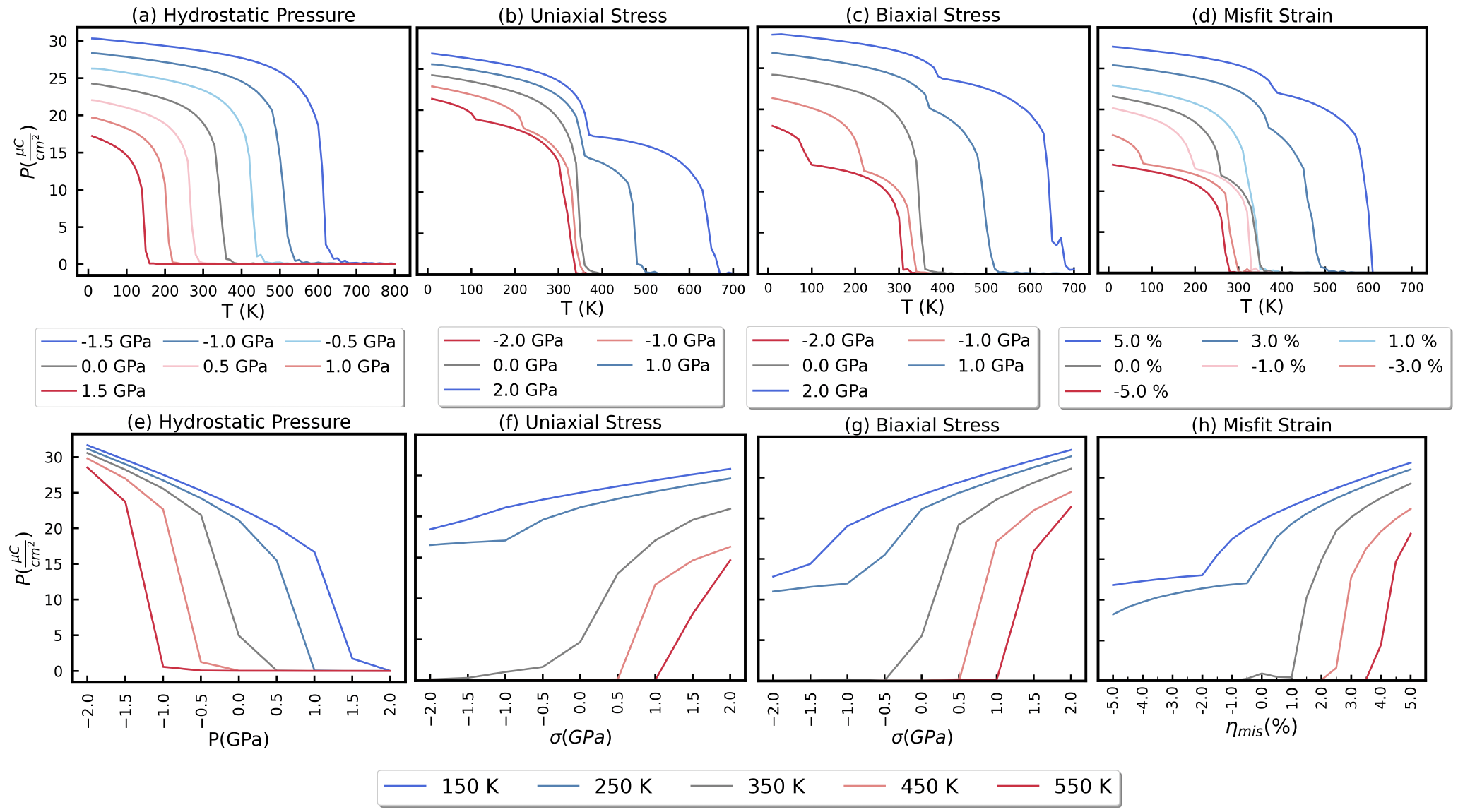}

\caption{ Polarization as a function of temperature for a few representatives of each type of the mechanical load (a)-(d); Polarization as a function of hydrostatic pressure (e); uniaxial stress (f); biaxial stress (g); and misfit strain (h) for a few representative temperatures.}
\label{figure1}
\end{figure}

Figure~\ref{figure1} gives the temperature evolution of polarization for a few representative stresses/strains for each type of mechanical load and compares them to the stress-free case. It reveals that additional phase transitions  can be induced in this material by external load.

\begin{figure}[h]
\centering
\includegraphics[width=1\textwidth]{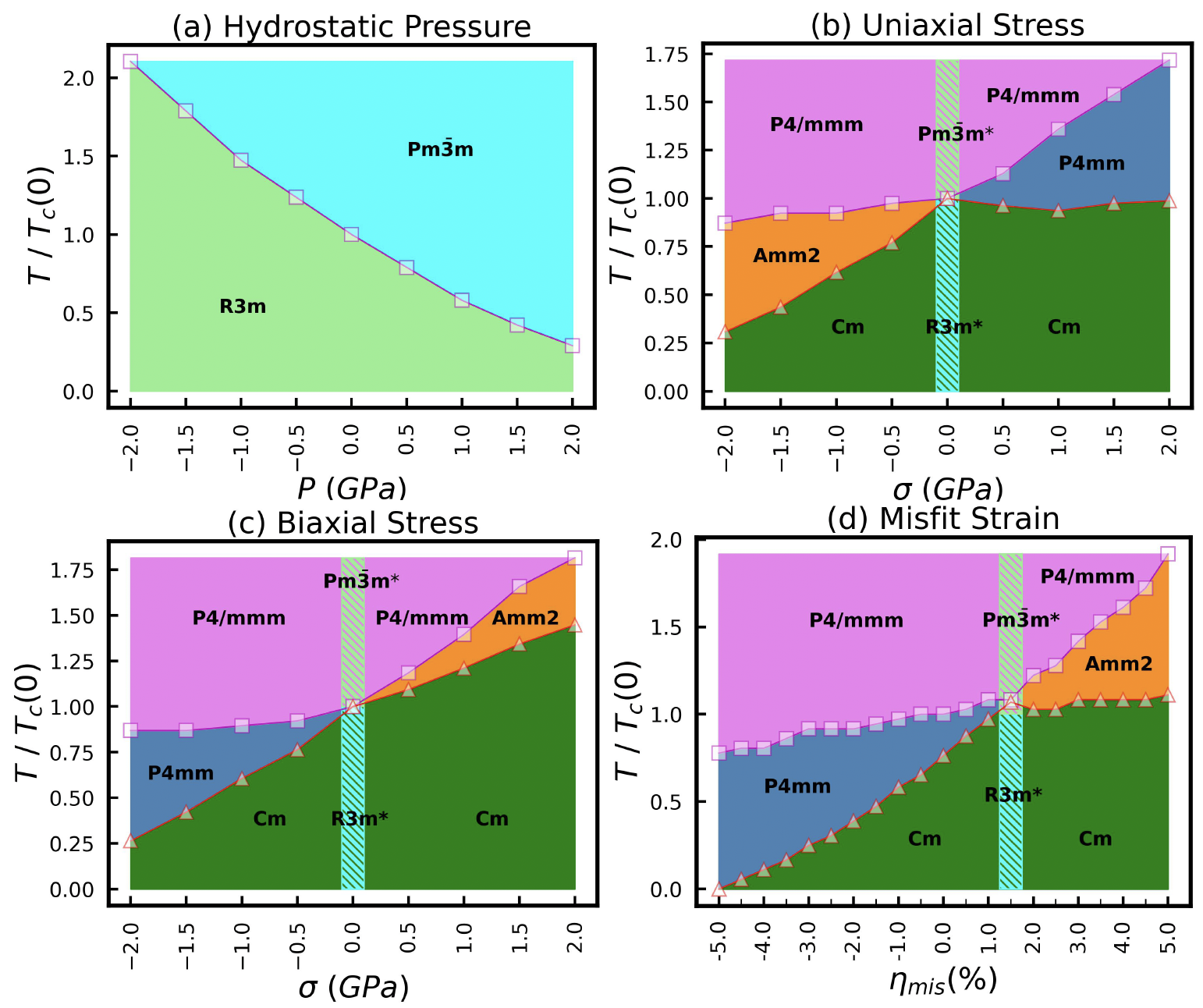}
\caption{ Phase diagrams for \cgb~predicted in computations: phase transition temperature as a function of hydrostatic pressure (a); uniaxial stress (b); biaxial stress (c); and misfit strain (d). * indicates phases in the absence of mechanical load.}
\label{figure2}
\end{figure}

For the case of hydrostatic pressure, we find that the positive (negative) pressure destabilizes (stabilizes) the ferroelectric phase with the Curie temperature change rate of $-172.3$~K/GPa, or $-0.45T_C(0)$~K/GPa. At room temperature, this yields a structural phase transition at a pressure of 0.94~GPa, which is within the range experimentally measured in Ref.~\onlinecite{PressureCGBexp} of 1.0(2)~GPa. 
The material undergoes a single phase transition, just as in the case of zero pressure. Figure~\ref{figure2}(a) shows the associated phase diagram. 

Computational data for uniaxial stress are presented in Fig.~\ref{figure1}(b) and reveal that the material now undergoes two phase transitions, as evident from the discontinuity in polarization. We find that tensile load increases Curie temperature, while compressive load decreases it. To the best of our knowledge, no other perovskite in bulk has shown this type of relationship before. For example, for bulk BaTiO$_3$, $T_C$ increases with both tensile and compressive stress/strain\cite{BTOPhaseDiagram}.
 
The associated phase diagram for uniaxial stress is given in Fig.~\ref{figure2}(b). For tensile uniaxial stress the material undergoes the following sequence of the phase transitions: paraelectric tetragonal P4/mmm to ferroelectric tetragonal P4mm to ferroelectric monoclinic Cm. Interestingly, we find that tensile stress drastically affects the Curie temperature, but the transition temperature between the two ferroelectric phases (P4mm and Cm) does not respond to stress. Under compressive uniaxial stress the material undergoes a phase transition from paraelectric P4/mmm phase to ferroelectric orthorhombic Amm2 and ferroelectric monoclinic Cm phases. The Curie temperature barely responds to the compressive stress, while the transition temperature between the two ferroelectric phases decreases as compressive stress increases in magnitude. In all cases the supercell remains in a monodomain phase. 

Figure~\ref{figure1}(c) displays temperature evolution of polarization in the presence of biaxial stress. We find that biaxial stress induces additional phase transitions and modifies the Curie temperature of the material. The associated phase diagram is given  in Fig.~\ref{figure2}(c). Similar to the case of uniaxial stress, we find that tensile load increases the Curie temperature, while compressive load suppresses it, but only very slightly. The phase diagram for biaxial stress is a mirror image of the one for uniaxial stress, which is rather intuitive, as compressive uniaxial stress expands the unit cell along the two perpendicular directions, somewhat mimicking biaxial tensile load.

The last type of the mechanical boundary condition we discuss is the misfit strain. The temperature evolution of polarization in the presence of misfit strain is given in Fig.~\ref{figure1}(d) and reveals that tensile strain stabilizes the ferroelectric phase and enhances the polarization, while compressive strain has an opposite effect. Once again we find that mechanical clamping induces an additional ferroelectric phase transition. The associated phase diagram is presented in Fig.~\ref{figure2}(d) and is qualitatively similar to the case of biaxial stress. The main differences are limited to the larger temperature range associated with stability of orthorhombic and tetragonal phases. 

Thus, our simulations predict that  most considered mechanical deformations are capable of inducing tetragonal, orthorhombic and monoclinic phases, none of which  are  available under unstressed conditions. The monoclinic phases are associated with a significant rotation of polarization from the polar axis of the rhombohedral phase, which could result in enhancement of piezoelectric response. In particular,  we find that the polarization rotates from the rhombohedral polar axes by 5.64\textdegree\  under 2.0~GPa tensile uniaxial stress and 10.17\textdegree\ under 2.0~GPa compressive uniaxial stress. Under biaxial stress, the polarization rotates from the rhombohedral polar axes by 6.18\textdegree\ under 2.0~GPa tensile biaxial stress and by 13.82\textdegree\ under 2.0~GPa compressive biaxial stress. For misfit strain,  the polarization rotates from the rhombohedral polar axes by 4.13\textdegree\ under 5.0\% tensile misfit strain and by 54.73\textdegree\ under 5.0\% compressive misfit strain. It should be noted that all the phases here are single domain. We did encounter polydomain phases in some of our annealing simulations. However, they are likely to be metastable as in all investigated cases a single domain phase can be stabilized using slower annealing  rate. 

Let us now investigate the origin of the peculiar features of the phase diagrams presented in Fig~\ref{figure2}, namely the decrease in the Curie temperature under compressive loads. We note that many phase diagrams for oxide ferroelectrics feature a V-shape\cite{BTOPhaseDiagram,McCash}, where the Curie temperature is enhanced by both tensile and compressive loads.  The effective Hamiltonian of Eq.~\ref{eq:1} can be minimized with respect to its independent variables following the approach of Ref.~\onlinecite{McCash}. We focus on uniaxial stress applied perpendicular to (100) plane, that is $\sigma_1$. Note, that the axes of our Cartesian coordinate system point along [100], [010] and [001] pseudocubic directions.  The local modes associated with the energy minima up to linear order in stress are 
\begin{equation}
    \begin{aligned}[c]
    u_x^2  =  u_{0,x}^2+a_x\sigma_1  \\
    u_y^2   =  u_{0,y}^2+a_y\sigma_1 
    \end{aligned}
    \label{eq:2}
\end{equation}
where the slopes are
\begin{equation}
   \begin{aligned}[c]
    a_x & =  \frac{ \frac{2\nu_t}{3\mu_t}(3\alpha'+\gamma') - \frac{\gamma'}{6}\frac{C}{B}}{2\gamma'(3\alpha'+\gamma')} \\ 
    \end{aligned}
    \qquad
    \begin{aligned}[c]
    a_y & =  \frac{-\frac{\nu_t}{3\mu_t}(3\alpha'+\gamma') - \frac{\gamma'}{6}\frac{C}{B} }{2\gamma'(3\alpha'+\gamma')} \\
 \end{aligned}
 \label{eq:3}
\end{equation}
where
\begin{equation}
   \begin{aligned}[c]
    C & =  B_{1xx} + 2B_{1yy} \\
    \nu_t & =  \frac{B_{1xx} - B_{1yy}}{2} \\
    \nu_r  & =  B_{5xz} \\
    \alpha' & =  \alpha - \frac{1}{24}(\frac{C^2}{B}+4\frac{\nu_t^2}{\mu_t})
    \end{aligned}
    \qquad
    \begin{aligned}[c]
    B & =  B_{11} + 2B_{12} \\
    \mu_t & =  \frac{B_{11}-B_{12}}{2} \\
    \mu_r & =  B_{44} \\
    \gamma' & =  \gamma + \frac{1}{2}(\frac{\nu_t^2}{\mu_t}-\frac{\nu_r^2}{\mu_r})
 \end{aligned}
    \qquad 
    \label{eq:4}
\end{equation}
and     
\begin{equation} 
u_{0,x}^2 = u_{0,y}^2 = -\frac{\kappa}{6\alpha'+2\gamma'} 
\label{eq:5} 
\end{equation}
 
\begin{figure}[h]
\centering
\includegraphics[width=1\textwidth]{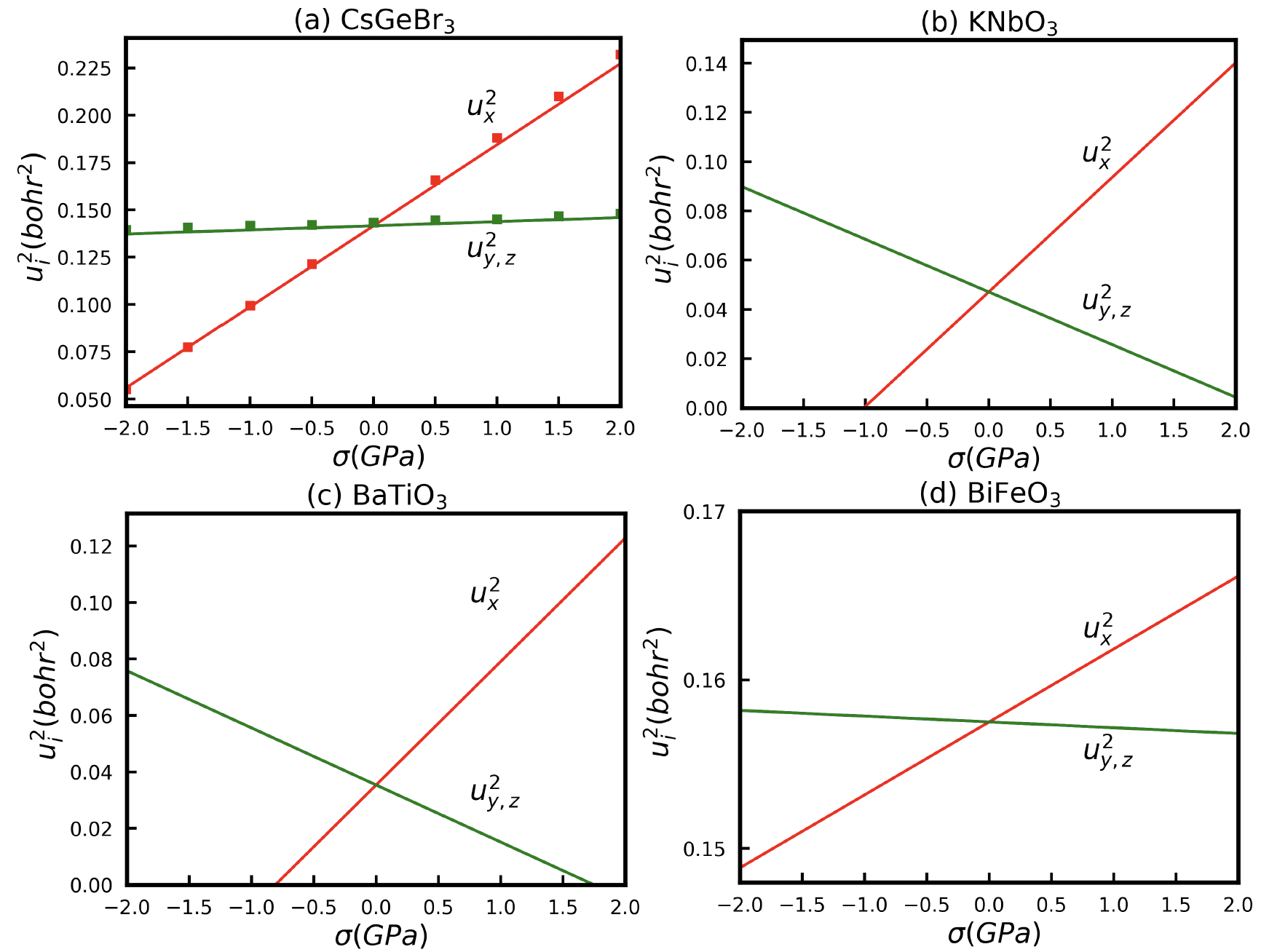}
\caption{ The dependencies of the components of the local mode squared, $u_i^2$, on the uniaxial stress obtained for 0~K from the analytical solution given in Eq.~\ref{eq:2} for different perovskite ferroelectrics, as specified in the titles. Symbols in (a) give the computational data from 10~K simulations.  }
\label{figure3}
\end{figure}

Parameters $ \kappa, \alpha, \gamma, B_{11}, B_{12}, B_{44}, B_{1xx}, B_{1yy}$, and $B_{5xz}$ are first-principles-based parameters of the effective Hamiltonian derived from DFT calculations in Ref.~\onlinecite{PolarAntipolar}. For \cgb   $\\a_x=$~1260~Bohr$^5$/Ha  and $a_y=$~64~Bohr$^5$/Ha, which results in dependencies given in Fig.~\ref{figure3}. The unique feature of these dependencies is the large difference between the slopes for the two components of the local mode. We find that the local mode (and therefore polarization) responds very strongly to the load along the stressed direction, while it barely responds in the transverse direction. This originates from the interplay between the $\frac{\nu_t}{3\mu_t}(3\alpha'+\gamma')$, which is a negative quantity, and $ \frac{\gamma'}{6}\frac{C}{B} $, which is positive, terms in Eq.~\ref{eq:3}. These  terms compete in $a_y$ and cooperate  in $a_x$, due to their opposing signs.

This  dependence of $u_x^2$ on stress results in the strong enhancement of $P_x$ under tensile load which leads to the  strong enhancement of $T_C$. On the other hand, under compressive load $P_x$ is greatly suppressed, while the $P_{y/z}$ are suppressed only slightly. So the paraelectric to ferroelectric phase transition is led now by the condensation of the latter components, which explains why the Curie temperature is slightly suppressed under compressive load. We note that the analytical predictions from Eqs.~\ref{eq:2}  closely match results from the simulations at the lowest temperature of 10~K (see Fig~\ref{figure3}(a)).

Our analytical model allows us to compare \cgb ~with oxide ferroelectrics which also undergo phase transitions into rhombohedral phase. Using the effective Hamiltonian parameters for BaTiO$_3$\cite{BTOKNOparams},  KNbO$_3$\cite{BTOKNOparams} and BiFeO$_3$\cite{BFOparams} we obtained the $u_i^2(\sigma)$ dependencies given in Fig.~\ref{figure3}(b)-(d). Interestingly, for all three of oxide ferroelectrics, the slopes for $u_x^2(\sigma)$ and $u_{y/z}^2(\sigma)$ have opposite sign, which results in the V-shape of the upper part of the phase diagram as previously discussed\cite{BTOPhaseDiagram,McCash}. So, the unique feature of \cgb ~is that its $T_C$ can be tuned both up and down by stresses and strains.

\section{Electromechanical coupling }
We now want to find out the isothermal response of polarization to pressure and stresses. Figure~\ref{figure1}(e)-(g) show the dependence of polarization on pressure/stress for a few representative temperatures. At the lowest temperature of 150~K, we find from Fig.~\ref{figure1}(a) that a pressure of 1.5~GPa destabilizes the ferroelectric phase. This critical pressure decreases as temperature increases.

Figure~\ref{figure1}(f) shows the isothermal polarization response to uniaxial stress and reveals its highly nonlinear nature. In the vicinity of the stress-induced phase transition, the slope of $P(\sigma)$ can reach very large values, promising outstanding piezoelectric response. Similar trends are found for biaxial stress [See Fig.~\ref{figure1}(g)]. Figure~\ref{figure1}(h) shows isothermal dependencies of polarization on misfit strain, reported for the sake of completeness, although these might be difficult to realize experimentally.

\begin{wrapfigure}{r}{0.5\textwidth}
  \begin{center}
    \includegraphics[width=0.48\textwidth]{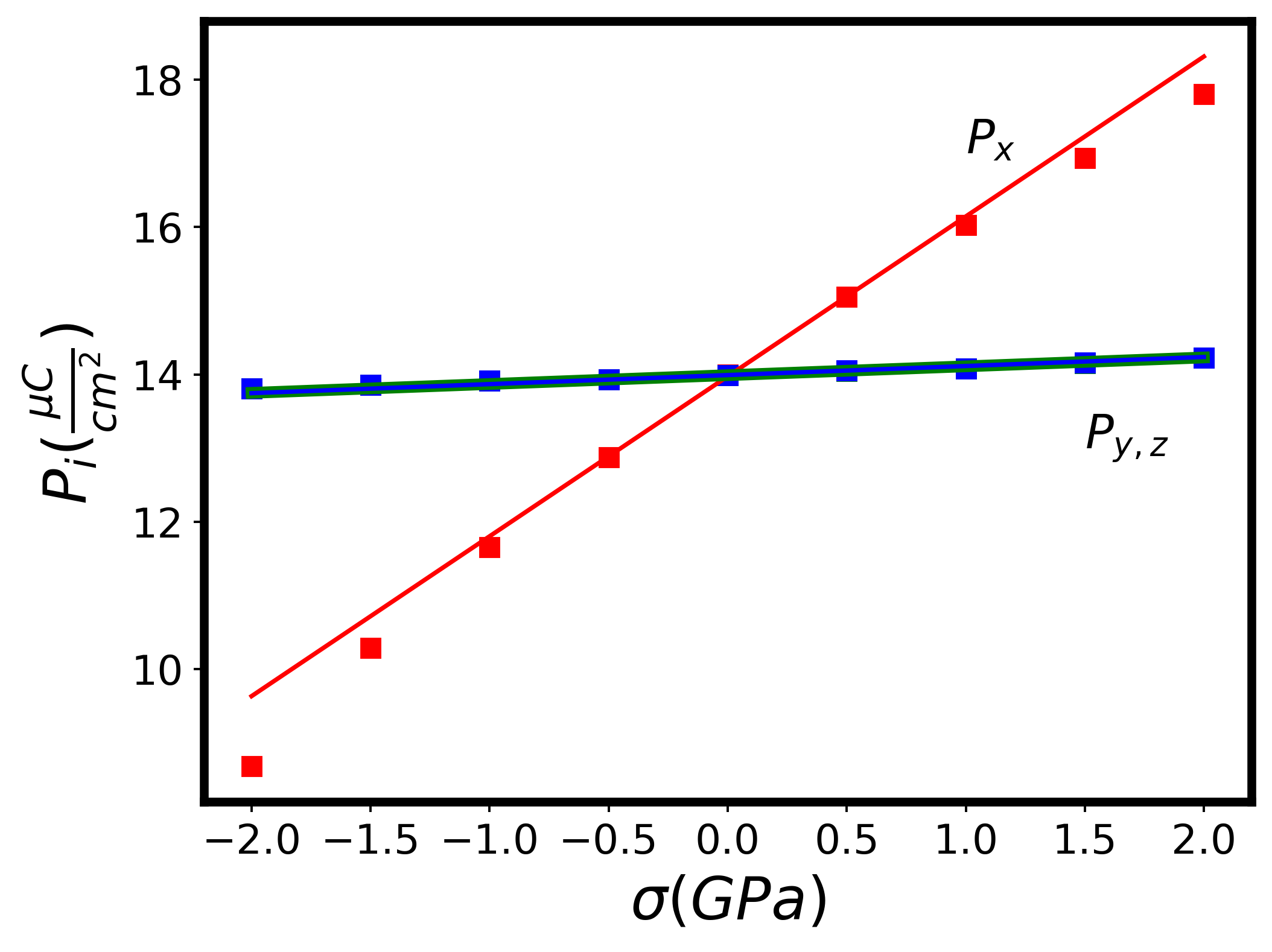}
  \end{center}
  \caption{Polarization components as functions of uniaxial stress computed at 10~K. Solid lines give the linear fit to the data.  }
  \label{figure4}
\end{wrapfigure}

To quantify electromechanical coupling in \cgb\ we compute the zero field slope of $P_i(\sigma)$ for the case of uniaxial stress using the data presented in Fig.~\ref{figure4}. The 10~K values (in pseudo cubic setting) are reported in Table~\ref{T1}. The $d_{11}$ coefficient is 21.7~pC/N, which is lower than in oxide ferroelectrics but significantly higher than in hybrid organic inorganic perovskites. 
For example, the longitudinal piezoelectric coefficient d$_{33}$ of typical oxide perovskites include 191~pC/N for BaTiO$_3$\cite{BTOpiezo}, 117~pC/N for PbTiO$_3$\cite{PropOfMat}, and 593~pC/N for PZT-5H\cite{PropOfMat}; while typical hybrid organic inorganic perovskite d$_{33}$ values include 17.1~pC/N for (BA)$_2$(MA)$_2$Sn$_3$Br$_{10}$\cite{OrgInorg} and 14~pC/N for MDABCO–NH$_4$I$_3$\cite{OrgInorg} and 1.18–21.12~pC/N in formates\cite{Formates}. Similarly, the isothermal dependencies of polarization on hydrostatic pressure can be used to compute  the hydrostatic piezoelectric coefficient, $d_h = 42.51$~pC/N, which is comparable to poled BaTiO$_3$, $d_h = 34.0$~pC/N\cite{BTOdh}.

\begin{table}[h]
\normalsize
\centering
\caption{\label{tab:properties}%
  Piezoelectric stress and strain coefficients ($e_{ij}$ and $d_{ij}$) in $C/m^2$ and pC/N, respectively, and elastic  constants ($C_{ij}$)  in kbar computed from  effective Hamiltonian ($H_{eff}$) and/or DFT. Coefficients $d_{ij}$ are reported in both standard (hexagonal) setting as well as the pseudocubic setting ($d_{ij}^{pc})$ to facilitate comparison between effective Hamiltonian and DFT predictions. }
\begin{tblr}
{Q[c,m,1.1cm]Q[c,m,1.2cm]Q[c,m,1.2cm]Q[c,m,1.1cm]Q[c,m,1.0cm]Q[c,m,1.0cm]Q[c,m,1.0cm]Q[c,m,1.0cm]Q[c,m,1.0cm]Q[c,m,1.0cm]Q[c,m,1.0cm]Q[c,m,1.0cm]}
\hline
\hline
$e_{15}$&  
$e_{22}$ &
$e_{31}$&
$e_{33}$&
$C_{11}$&
$C_{12}$&
$C_{13}$ &
$C_{14}$ &
$C_{33}$ &
$C_{44}$ &
$C_{65}$ &
$C_{66}$ \\
\hline
-0.23 & 0.10 & -0.27 & -0.52 & 222.15 & 61.30 & 74.44 & 31.51 
& 155.29 & 94.14 & 31.50 & 80.01 \\
\hline
\hline
$d_{15}$ & $d_{22}$ & $d_{31}$ & $d_{33}$ & $d_{11}^{pc}$ & $d_{12}^{pc}$ & $d_{14}^{pc}$  & $d_{15}^{pc}$ & $d_{11}^{pc}$  \scriptsize{(Heff)}  & $d_{12}^{pc}$ \scriptsize{(Heff)} & $d_{14}^{pc}$ \scriptsize{(Heff)} & $ d_{15}^{pc} $ \scriptsize{(Heff)}
\\
\hline
-23.39 & 1.93 & -0.81 & -33.05 & 16.61 & 1.55 & 5.45 & 16.72 
& 21.70 & 1.23 & - & - \\

\hline

\end{tblr}
\label{T1}
\end{table}

\section{Insights from DFT simulations}

In this section we take advantage of DFT simulations to both  validate our effective Hamiltonian predictions and reach additional insights into the electromehanical properties of \cgb. We begin by validating the phases predicted by the effective Hamiltonian under misfit strain. For that we pick one point from each unique area of the temperature-misfit  strain phase diagram, Fig.~\ref{figure2}(d), and use the associated structure to initialize DFT simulations. Technically, the average values of local mode and lattice constants are used to  reconstruct atomic structure from effective Hamiltonian degrees of freedom. We keep the in-plane lattice vectors constrained to the values reconstructed from the  effective Hamiltonian simulations but  allow for full ionic relaxation and relaxation of the remaining lattice vector. All the phases predicted by the effective Hamiltonian calculations retained their space groups. Table~\ref{T2}  summarizes these phases, conditions at which they were obtained, their structural parameters as well as their energies and enthalpies. The structural files are also provided in Ref. \onlinecite{GITHUB}. The energies of most phases are significantly higher than the ground state (R3m), which is due to the mechanical constraints and finite temperatures. Enthalpies are reported with respect to ground state (R3m) for the phases harvested below the Curie point and with respect to cubic phase (Pm$\bar{3}$m) for  the phases obtained above the Curie point. In all cases the relative enthalpies are negative, validating the stability of the predicted phases under chosen conditions. 

\begin{table}[h]
\normalsize
\centering
\caption{\label{tab:properties}%
  Structural parameters (a, b, c, $\alpha$, $\beta$, and $\gamma$), relative energies ($\Delta E$) and enthalpies ($\Delta H$) computed from DFT for different phases obtained from the temperature-misfit strain phase diagram points ($\eta_{mis}$, $T$).  Enthalpies are reported relative to cubic phase (Pm$\bar{3}$m) for phases above T$_C$ and relative to the ground state (R3m) for phases below T$_C$.  Structures are symmetrized using the FINDSYM software package\cite{FINDSYM,FINDSYM2}.}
\begin{tblr}
{Q[c,m,1.0cm]Q[c,m,1.2cm]Q[c,m,1.2cm]Q[c,m,1.0cm]Q[c,m,1.0cm]Q[c,m,1.0cm]Q[c,m,1.0cm]Q[c,m,1.0cm]Q[c,m,1.0cm]Q[c,m,1.0cm]Q[c,m,1.0cm]Q[c,m,1.0cm]}
\hline
\hline
Space Group& 
$T$ \small{(K)} &
\textit{$\eta_{mis}$}&
a (\AA)&
b (\AA)&
c (\AA)&
\textit{$\alpha$} &
\textit{$\beta$} &
\textit{$\gamma$} &
\normalsize{\textit{$\Delta$}E} \tiny{(meV/f.u.)} &
\normalsize{\textit{$\Delta$}H$_{cubic}$} \tiny{(meV/f.u.)}&
\normalsize{\textit{$\Delta$}H$_{R3m}$} \tiny{(meV/f.u.)}\\ 
\hline
\scriptsize{P4/mmm} & 600 &  -4.0\% & 5.32 & 5.32 & 5.64 & 90.0 & 90.0 & 90.0 & 235 & -146 & - \\
\scriptsize{P4/mmm} & 600 & +2.5\% & 5.68 & 5.68 & 5.49 & 90.0 & 90.0 & 90.0 & 149 & -33 & -  \\
\scriptsize{P4mm} & 170 & -4.0\% & 5.32 & 5.32 & 5.79 & 90.0 & 90.0 & 90.0 & 200 & - & -220  \\
\scriptsize{Amm2} & 500 & +5.0\% & 5.53 & 8.23 & 8.23 & 90.0 & 90.0 & 90.0 & 58 & -119 & -  \\
\scriptsize{Cm} & 10 & -4.0\% & 7.52 & 7.52 & 5.79 & 90.0 & 90.1 & 90.0 & 199 & - & -69  \\
\scriptsize{Cm} & 10 & +5.0\% & 8.23 & 8.23 & 5.64 & 90.0 & 90.4 & 90.0 & 32 & - & -32  \\
\hline

\end{tblr}
\label{T2}
\end{table}

We next validate and augment our effective Hamiltonian predictions of the piezoelectric response using DFT calculations. Table~\ref{T1} reports the piezoelectric stress and strain tensor components computed from DFT.  For the strain coefficients, we report the values in the hexagonal setting, as well as the pseudocubic setting to facilitate  comparison with the effective Hamiltonian predictions. Note that we cannot convert the effective Hamiltonian piezoelectric tensor to hexagonal setting, as we cannot resolve all of its components from uniaxial stress application.  We find good agreement between effective Hamiltonian and DFT predictions for the piezoelectric response. DFT predicts the largest piezoelectric coefficient is $d_{33}=33.05$~pC/N, which is comparable to rhombohedral BiFeO$_3$: $d_{33} = 27-45$~pC/N\cite{BFOpiezo,BFOpolar2}.

\section{Conclusions}

In summary, we used a combination of finite-temperature effective Hamiltonian simulations and zero Kelvin DFT simulations to study ferroelectric phases and phase transitions in the Ge-based ferroelectric halide perovskite \cgb. We found that hydrostatic pressure has a significant effect on the Curie temperature and decreases it at the rate $-0.45$~$T_C$(0)~K/GPa, where $T_C$(0) is the zero pressure Curie temperature.  However, the phase transition sequence remains the same as for the unstressed case, namely paraelectric Pm$\bar{3}$m to ferroelectric R3m. Application of uniaxial or biaxial stress, as well as misfit strain, induces a completely different sequence of phase transitions.  Under biaxial tensile stress/strain, the material undergoes a phase transition from paraelectric tetragonal phase to ferroelectric orthorhombic phase, followed by another ferroelectric phase transition to monoclinic phase. Such tensile load dramatically enhances the Curie  temperature, up to 2 times under the largest investigated strain of 5\%. Biaxial compressive stress/strain slightly suppresses the Curie point. The material undergoes a sequence of phase transitions from paraelectric tetragonal phase to ferroelectric tetragonal phase to ferroelectric monoclinic phase. The ferroelectric-ferroelectric phase transition temperature is highly sensitive to the load. Uniaxial stress has an opposite effect on the phase transitions and the associated transition temperatures in comparison with the biaxial stress, which could be justified by the deformations in the unit cell produced by such loads. This behavior originates from the interplay between the material parameters that describe electrostriction, elastic properties, and the strength of the ferroelectric instability.
 The zero field piezoelectric constants were found to be in the range 0.8-33.1~pC/N, which is comparable to BiFeO$_3$\cite{BFOpiezo}. Strong nonlinearity in the response of polarization to stress has been predicted,  which could result in outstanding energy converting properties and find applications in energy converting devices. We believe that our work lays a foundation for phase and ferroelectricity manipulation in lead-free perovskite materials, which could lead to the experimental discoveries  of these phases and potential use of Ge-based ferroelectrics for environmentally friendly lead-free energy converting technologies.

\section{Acknowledgement}

This work was supported by the U.S.
Department of Energy, Office of Basic Energy Sciences, Division of Materials Sciences and Engineering under Grant No. DE-SC0005245.  Computational support was provided by the National Energy Research Scientific Computing Center (NERSC), a U.S. Department of Energy, Office of Science User Facility located at Lawrence Berkeley National Laboratory, operated under Contract No. DE-AC02-05CH11231 using NERSC award BES-ERCAP-0025236.

\bibliography{main}

\end{document}